\documentclass[aps,prl,reprint,superscriptaddress,noeprint]{revtex4-1}

\usepackage{listings}
\usepackage{graphicx}
\usepackage{dcolumn}
\usepackage{bm}
\usepackage{xcolor}
\usepackage{changes}
\usepackage{comment}
\usepackage[mathlines]{lineno}
\modulolinenumbers[5]

\begin{document}

\title[Free-Form Dark Hollow beams]{Geometric-Phase Waveplates for Free-Form Dark Hollow beams} 

\author{Bruno Piccirillo}
\email{bruno.piccirillo@unina.it.}
 \affiliation{Department of Physics ``E. Pancini'', Universit\`a  di Napoli Federico II, Complesso Universitario MSA, via Cintia, 80126, Fuorigrotta-Napoli, Italy
}%

\author{Ester Piedipalumbo}%
\email{ester.piedipalumbo@unina.it.}
\affiliation{Department of Physics ``E. Pancini'', Universit\`a  di Napoli Federico II, Complesso Universitario MSA, via Cintia, 80126, Fuorigrotta-Napoli, Italy
}%

\affiliation{INFN-Sezione di Napoli, Complesso Universitario MSA, via Cintia, 80126, Fuorigrotta-Napoli, Italy}

\author{Enrico Santamato}
\affiliation{Department of Physics ``E. Pancini'', Universit\`a  di Napoli Federico II, Complesso Universitario MSA, via Cintia, 80126, Fuorigrotta-Napoli, Italy
}%

\date{\today}

\begin{abstract}

We demonstrate the possibility to create optical beams with phase singularities engraved into exotic intensity landscapes imitating the shapes of a large variety of diverse plane curves. To achieve this aim, we have developed a method for directly encoding the geometric properties of some selected curve into a single azimuthal phase factor without passing through indirect encryption methods based on lengthy numerical procedures. The outcome is utilized to mould the optic axis distribution of a liquid-crystal-based inhomogeneous waveplate. The latter is finally used to sculpt the wavefront of an input optical gaussian beam via Pancharatnam-Berry phase.

\end{abstract}

\maketitle

\section{Introduction}
Light sculpting has gained increasing importance in both fundamental and applied optics~\citep{Rubinsztein_Dunlop_2016}. Engraving singularities in optical beams, in particular, has paved the way for multiple applications in both classical and quantum optics, most of which related to the angular momentum of light. Singular Optics has gradually become an independent research field and now aspires to become a fundamental cornerstone of modern photonics. Optical singular beams have proven to be invaluable for non-contact manipulation over micro- and nanoscale~\citep{TAYLOR2015669,DHOLAKIA2011335}  -- which has enormous implications on modern nanophysics, crystal growth, metamaterials, to give just a few examples. Furthermore, the infinite dimensionality of the orbital angular momentum (OAM) space has paved the way for increasing data capacity of both free-space and fiber-optic communications~\citep{Bozinovic20131545} and for developing novel efficient protocols for classical~\citep{Willner:15} as well as quantum information processing~\citep{Mair2001313,Molina-Terriza2007305,Peacock20161152}. No less important, optical singularities have been successfully utilized for super-resolution imaging~\citep{Maurer2010912,daguanno2008043825}, on-chip optical switching~\citep{C1NR11406A,Boriskina:11,doi:10.1021/nl903380j}, advanced microscopy~\citep{Hell:94,PhysRevLett.86.5251} and material machining~\citep{Nivas201742142,Meier2007,doi:10.1063/1.4793668}. 

Needless to say the great potential of singular optics -- and, more generally, of sculpted light -- has been progressively  unlocked in time, through the development of increasingly efficient and versatile tools for shaping the optical wavefronts. The most prominent technologies currently available for shaping spatial modes are computer generated holograms (CGHs) displayed on spatial light modulators (SLMs) -- based on dynamic phase control -- and Pancharatnam-Berry phase Optical Element (PBOEs) or geometric phase. Indeed, several methods are nowadays available to fabricate geometric-phase optical elements for wavefront-shaping, ranging from subwavelength metal stripe space-variant gratings~\citep{bomzon01}, to multilayer plasmonic metasurfaces~\citep{Liu:16} and Spatially Varying Axis Plates (SVAPs) based on liquid crystals~\citep{Kim:15,doi:10.1002/adom.201800961, piccirillo10, 10.1117/12.2078372, Piccirillo:17, Aleman-Castaneda:19}.

In the present paper, we introduce a method for designing SVAPs enabling to generate scalar optical beams with nonlinear azimuthal phase structures giving birth to phase singularities engraved within non-cylindrically symmetric intensity profiles. Indeed, the cylindrical symmetry typical of the intensity profile of helical beams springs from their linear azimuthal phase profile, $e^{{\rm i}\,\ell \phi}$. Helical beams have helical wavefronts -- hence the name --  and carry an OAM of $\hbar \ell$ per photon, $\ell$ being an integer number and $\phi$ the azimuthal polar angle around the beam propagation direction. There are multiple families of helical beams differing for their radial dependences. Well-known examples are Laguerre-Gaussian (LG) beams~\citep{allen92,allen}, Bessel and Bessel-Gaussian (BG) beams~\citep{PhysRevA.71.033411} and the wider class of Hypergeometric-Gaussian (HyG) beams~\citep{karimi08}, to name just a few. A helical beam with an azimuthal index $\ell$ has an $\ell$-fold rotational symmetry and its OAM spectrum accordingly includes only the component $\ell$. Denoted as $\bm k$ the light beam wavevector, the azimuthal component of the linear momentum is $\hbar k_{\phi}$ per photon: it does not depend on $\phi$, but only on the distance from the beam axis. The energy flux is therefore rotationally invariant around the beam axis, yielding the well-known cylindrically-symmetric doughnut-shaped profile. An azimuthally nonuniform $k_{\phi}$, in contrast, will break such symmetry and will give birth to an optical wavefront with a nonuniform helical phase structure, which will result, in its turn, into a non-cylindrically symmetric intensity profile. An OAM spectrum will broaden as a consequence of such symmetry breaking.

To impart a nonlinear azimuthal structure, we have developed a phase design method aimed at encoding the geometric properties of some plane curve, in order to create an intensity profile imitating the shape of the curve. We presently demonstrate that such an approach enables to directly determine the phase profile required to reshape the intensity profile of a light beam, as well as its OAM spectrum according to one's wishes. Here, in fact, we avoid passing through indirect methods for encoding amplitude and phase of the target field into a single phase function~\citep{Bolduc:13}. Though, the price to be payed is that only some features of the intensity profile and of the OAM spectrum will be precisely determined. Despite these apparent limitations, our method spontaneously leads us to introduce the concept of dark hollow beams with tailored intensity profiles or ``Free-Form Dark-Hollow'' (FFDH) Beams. A detailed study of the optical properties of FFDH beams will be reported elsewhere. Here we focus the attention on the generation of such beams by using the aforementioned  SVAPs, of which q-plates~\citep{marrucci06} are probably the most famous examples. Liquid-crystals based SVAPs combine high conversion efficiency with exceptional manageability for overall high performance. 
 Our SVAPs were fabricated adopting a ``direct-write approach", as defined in Ref.~\citep{Kim:15}. However, we would like to emphasize that our focus is presently on the method worked out to determine the transmittance phase function. Specifically, an arbitrary superposition of azimuthal modes amounts to a complex function of $\phi$ with both an amplitude and a phase, i.e.
\begin{equation}\label{eq:target}
 \sum_\ell c_\ell {\rm e}^{{\rm i}\ell \phi}=A(\phi){\rm e}^{{\rm i} \Psi(\phi)}.
\end{equation}
Several approaches mostly based on Gerchberg-Saxton algorithm are usually adopted to obtain a pure phase function providing an acceptable approximation for Eq.~(\ref{eq:target})~\citep{lu:16}. In what follows, we describe a method to directly generate a dark hollow beam in which the shape of the dark zone is basically inherited from the shape of a selected plane curve. This is achieved without recurring to inverse algorithms such as those mentioned above. They can be proved to be promising devices of potential interest for multiple applications ranging from super-resolution microscopy, to directional selective trapping~\citep{Porfirev:15}, as well as material processing, optical coronagraphy, not to mention the applications to classical and quantum communications~\citep{Piccirillo:19b,Rubano:19}. As an example, we consider the case of Stimulated Emission Depletion (STED) microscopy, in which super resolution is achieved by the selective deactivation of fluorophores through an excitation beam filling the internal zone of doughnut-shaped de-excitation spot. Replacing the doughnut with an FFDH beam, the illumination area would acquire a non-circular shape, suitable for optimally send photons to zones where they are really required and/or to prevent them from damaging the surrounding areas.
\section{Free-Form Azimuthal Phase Shaping}
The question arises as to what extent the transverse intensity profiles or the OAM spectrum of a light beam can be moulded by manipulating a purely azimuthal phase factor $e^{{\rm i}\psi(\phi)}$, $\psi(\phi)$ being an arbitrary function of the azimuthal coordinate $\phi$. Such a phase factor does not enable to explore all the possible field distributions, neither approximately, since $\psi$ is assumed independent of the distance $r$ from the beam axis~\citep{Bolduc:13,Forbes:16}. As above mentioned, in this work, we aim at introducing a \textit{toy}-method based on geometric intuition to determine the most appropriate azimuthal phase factor $e^{{\rm i}\psi(\phi)}$ required to generate dark hollow beams having arbitrary shapes or, as we have baptized them, FFDH beams. To this purpose, we need a ``dough cutter'' for partitioning the plane around the beam axis into a number of sectors -- ``slicing the doughnut'' -- and then distribute the transverse intensity of light among the several sectors according to one's wishes and necessities. Moulding the intensity of light within each sector is necessary for tailoring the boundaries of the dark region around the axis -- ``shaping the hole of the doughnut''. The portions of light within different sectors can be disconnected from each other or not. Metaphors aside, our ``dough cutter'' is the azimuthal component $\hbar k_{\phi}(\phi)$ of the photon linear momentum as a function of $\phi$, i.e.
\begin{equation}\label{eq:kphi}
k_{\phi}(\phi)=\frac{1}{r}\frac{d \psi(\phi)}{d \phi}.
\end{equation}
Assuming $\psi({\phi})$ proportional to the orientation angle $\Theta(\phi)$ of the unit normal to some plane curve $\gamma$ described by $\phi$-dependent parametric equations, then all the relevant features of $k_{\phi}(\phi)$ can be gathered from the rotational symmetry properties of $\gamma$ and from the local radius of curvature -- the latter being related to both $k_\phi$ and its derivative. Such a geometric approach has the advantage that $\Theta(\phi)$ -- and therefore the plane curve it comes from -- needs not to be determined, on a case-by-case basis, as a solution of an inverse problem. Rather, it can be helpful using a representation of the curve in polar coordinates, with some free parameters that can be tuned to match as much as possible the target intensity profile.

\subsection{Curve selection} 
Multiple choices are available. Good options are Lam\'e curves or their generalizations. A Lam\'e Curve, also known as a superellipse~\citep{Hazewinkel01}, is a closed curve retaining the geometric properties of semi-major axis and semi-minor axis, typical of an ellipse, but with a different shape.
In polar coordinates it is described by the equation
\begin{equation}
\left(a \cos \phi \right)^{\frac{n}{n-1}}+ \left( b \sin \phi \right)^{\frac{n}{n-1}}= \rho(\phi)^{\frac{n}{n-1}}\,,
\label{lame}
\end{equation}
where $a$, $b$, and $n$ are positive reals.

In 2003, J. Gielis introduced a single parametric equation -- dubbed the ``superformula'' -- describing multiple plane curves, of the most varied kinds, to study forms in plants and other living organisms~\citep{gielis2003a}. The mathematical expression of the superformula, in polar coordinates, is
\begin{equation}\label{eq:rho}
\rho(\phi)=
\left( \left| \frac{\cos{\frac{m \phi}{4}}} {a} \right |^{n_2} + \left| \frac{\sin{\frac{m \phi}{4}}}{b}  \right |^{n_3} \right)^{-\frac{1}{n_1}},
\end{equation}
where $\rho$ is the distance of a point of the curve $\gamma$ from the origin of the coordinate system as a function of the azimuthal angle $\phi$, $m$ is an integer number, $n_1$, $n_2$ and $n_3$ are three integers controlling its local radius of curvature and, finally, the positive real numbers $a$ and $b$ parameterize the radii of the circumferences respectively inscribed and circumscribed to the curve $\gamma$. For even $m=2\,k$,  Eq.~(\ref{eq:rho}) describes a curve $\gamma_{2\,k}$ closing over the interval $\left[0,2\pi \right)$. $\gamma_{2\,k}$ is rotationally symmetric by an angle $2\pi/k$. For odd $m=2\,k+1$, $\gamma_{2\,k+1}$ closes over the interval $\left[0,4\pi \right)$. When $a = b$ and $n_1=n_2$, $\gamma_m$ exhibits an $m$-fold rotational symmetry $C_m$. As varying all the free parameters in Eq.~(\ref{eq:rho}), the generated curves can be deeply diverse. No doubt the curves could be grouped according to a criterium based on the order of the their rotational symmetry. For $m=4$, $a=b$ and $n_2=n_3>2$, for instance, the superformula simply returns the superellipses first introduced by G. Lam\'e in 1818~\citep{Hazewinkel01}. For fixed values of $m$, $a$ and $b$, however, the signs and the absolute values of $n_1$, $n_2$ and $n_3$ can dramatically change the topological properties of the curves. Besides, a peculiar feature of the superformula is the fact that, independently of $m$, when $n_2=n_3=2$, it always degenerates into a circumference when $a=b$, or into an ellipse otherwise. Here, we are not interested in the mathematical peculiarities of the superformula, but rather in taking advantage of its ``shape-shifter'' capabilities.

\textit{Encrypting the geometrical properties of the selected curves into the optical phase.} 
Be $\gamma(a,b,m,n_1,n_2,n_3)$ the curve described by the superformula for some values of the free parameters.  The normal unit vector $\bm{n}=\left(n_x,n_y\right)$ of the curve is given by 
\begin{equation}\label{eq:normal}
\left(n_x + {\rm i} \, n_y\right)^2=\frac{\rho(\phi)-{\rm i}\,\dot{\rho}(\phi)}{\rho(\phi)+{\rm i}\,\dot{\rho}(\phi)} e^{2\,{\rm i}\phi},
\end{equation}
where $\dot{\rho}$ is the derivative of $\rho$ with respect to $\phi$. Denoted as $\Theta(\phi)$ the angle that $\bm{n}$ forms with the $x-$axis, we set the optical phase $\psi(\phi)$ to be
\begin{equation}\label{eq:superphase}
\psi(\phi)=2 \, \Theta(\phi; a,b,m,n_1,n_2,n_3).
\end{equation}
Consequently, as varying the free parameters in Eq.~(\ref{eq:rho}), multiple phase profiles can be designed and FFDHs accordingly generated. The realized phase profiles exhibit a modulation having the same symmetry properties as the curve $\gamma$. In the following we show that the $m$-fold symmetry characterizing the phase modulation affects also the intensity profile of the generated beam. Light intensity, indeed, is expected to be equally partitioned among the $m$ equally spaced sectors of the phase profile.

\begin{figure*}[h!]
\begin{center}
\includegraphics[width=10cm]{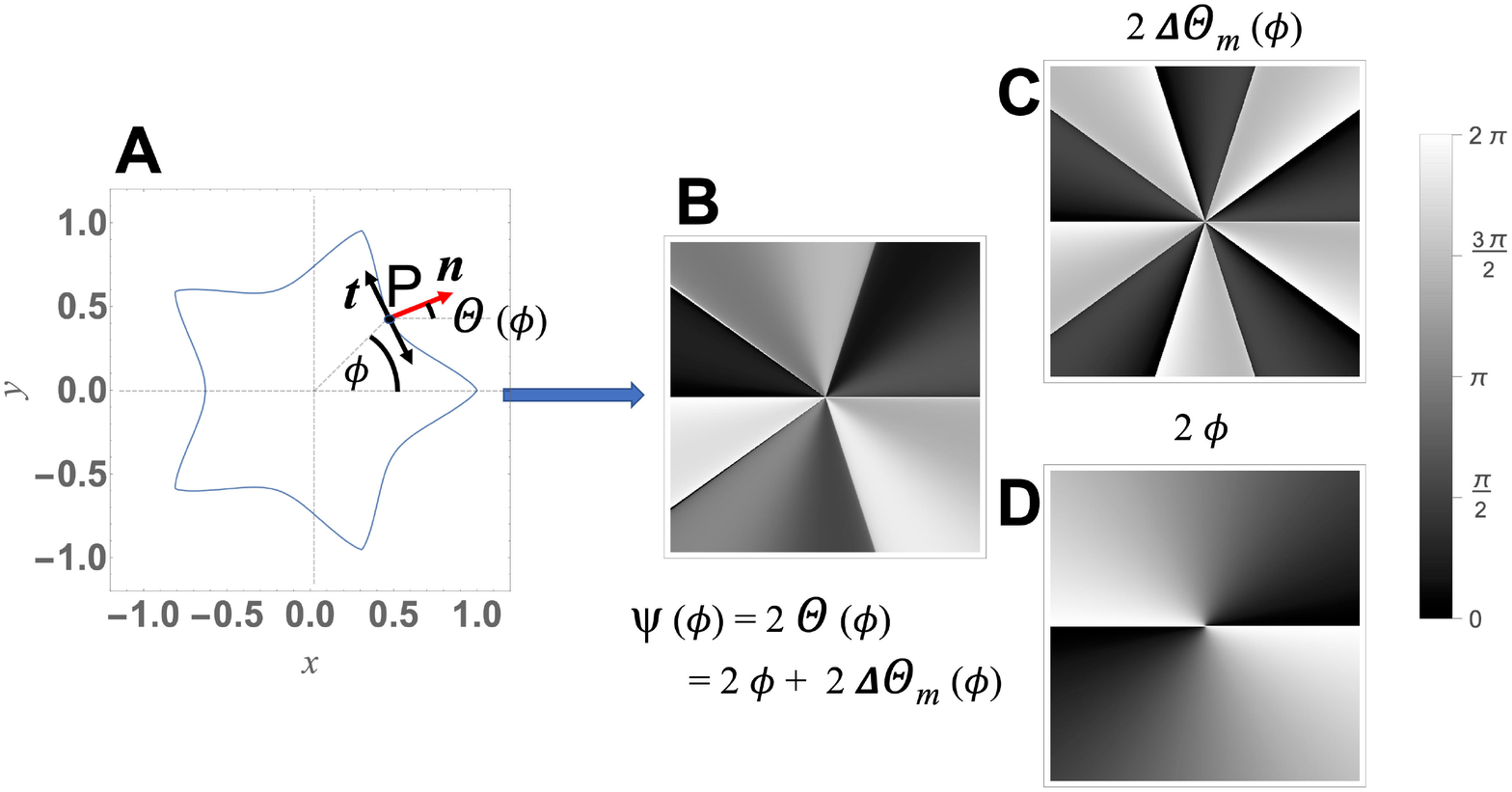}
\end{center}
\caption{Schematic of the encryption procedure of the symmetry properties of a plane curve into the azimuthal phase of a light beam. In inset {\bf A}, as an example, we show $\bm{t}$ and $\bm{n}$, i.e. the tangent and the normal unit vectors to the curve in the point $P$ respectively. $\bm n$ forms the angle $\Theta(\phi)$ with the horizontal axis. In inset {\bf B}, we show the transverse phase profile $\psi(\phi)=2 \, \Theta(\phi; 1, 1, 5, 1/2, 4/3, 4/3)$ (Eq.~(\ref{eq:superphase})). The latter can be regarded as the superposition of the phase modulation $2 \, \Delta \Theta_m(\phi)$ (inset {\bf C})  and the helical phase profile $2\,\phi$ (inset {\bf D}).}\label{fig:Fig1}
\end{figure*}

\begin{figure*}[h!]
\begin{center}
\includegraphics[width=10cm]{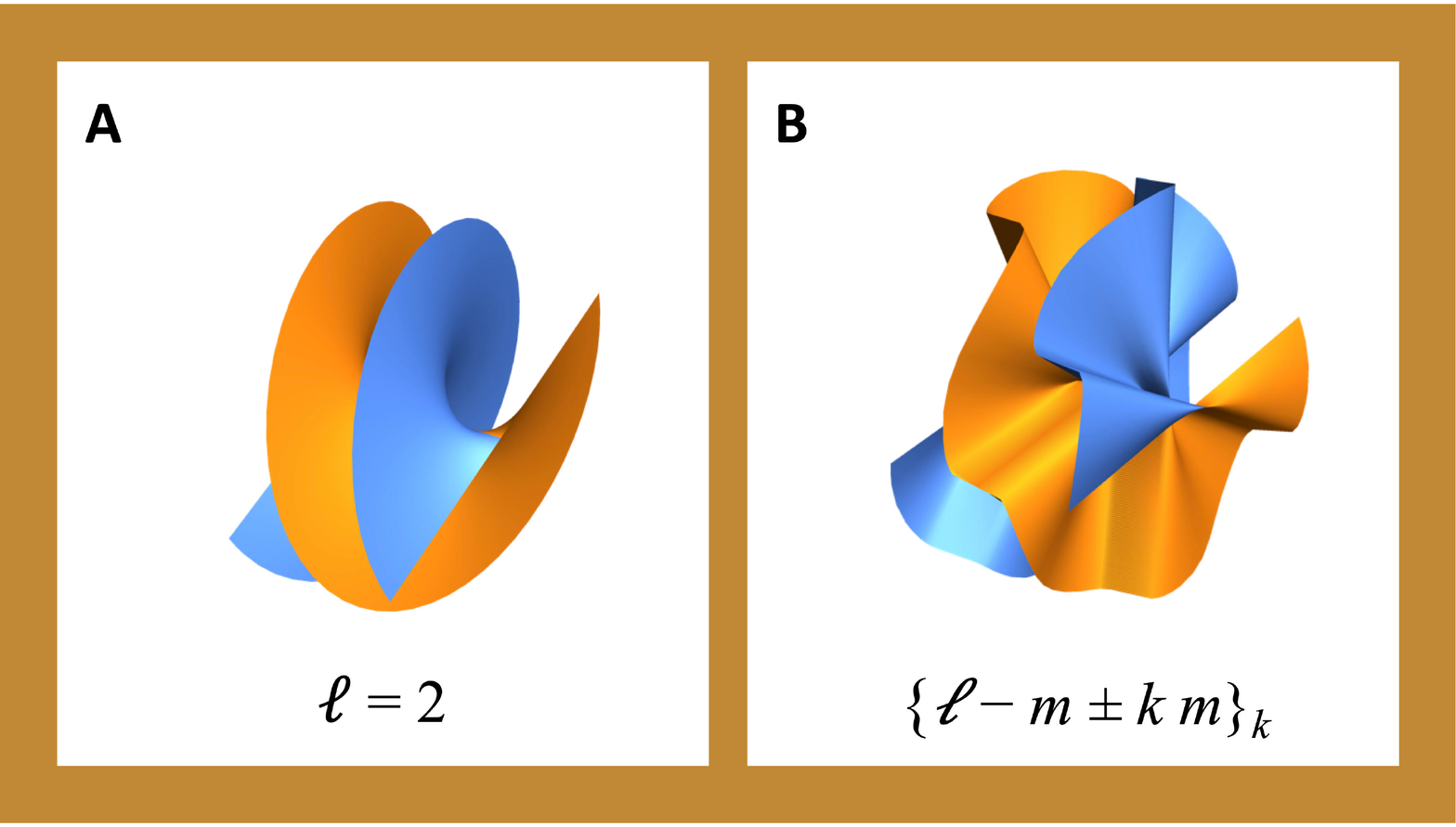}
\end{center}
\caption{Helical wavefronts corresponding to a circumference (inset {\bf A}) and to a the curve represented in Fig.~\ref{fig:Fig1}~{\bf A} (inset {\bf B}).}\label{fig:Fig2}
\end{figure*}

In Fig.~\ref{fig:Fig1}, this geometry-to-phase transfer procedure is sketched in the case $a=b=1$, $m=5$, $n_1=1/2$ and $n_2=n_3=4/3$. The rippled helical wavefront arising from Eq.~(\ref{eq:superphase}) is shown in Fig.~\ref{fig:Fig2} ({\bf B})) for the same values of the parameters and is compared to the smooth helical wavefront corresponding to a doughnut beam with $\ell =2$ Fig.~\ref{fig:Fig2} ({\bf A})). The latter can be easily shown to come from a circumference.

\begin{figure*}[h!]
\begin{center}
\includegraphics[width=10cm]{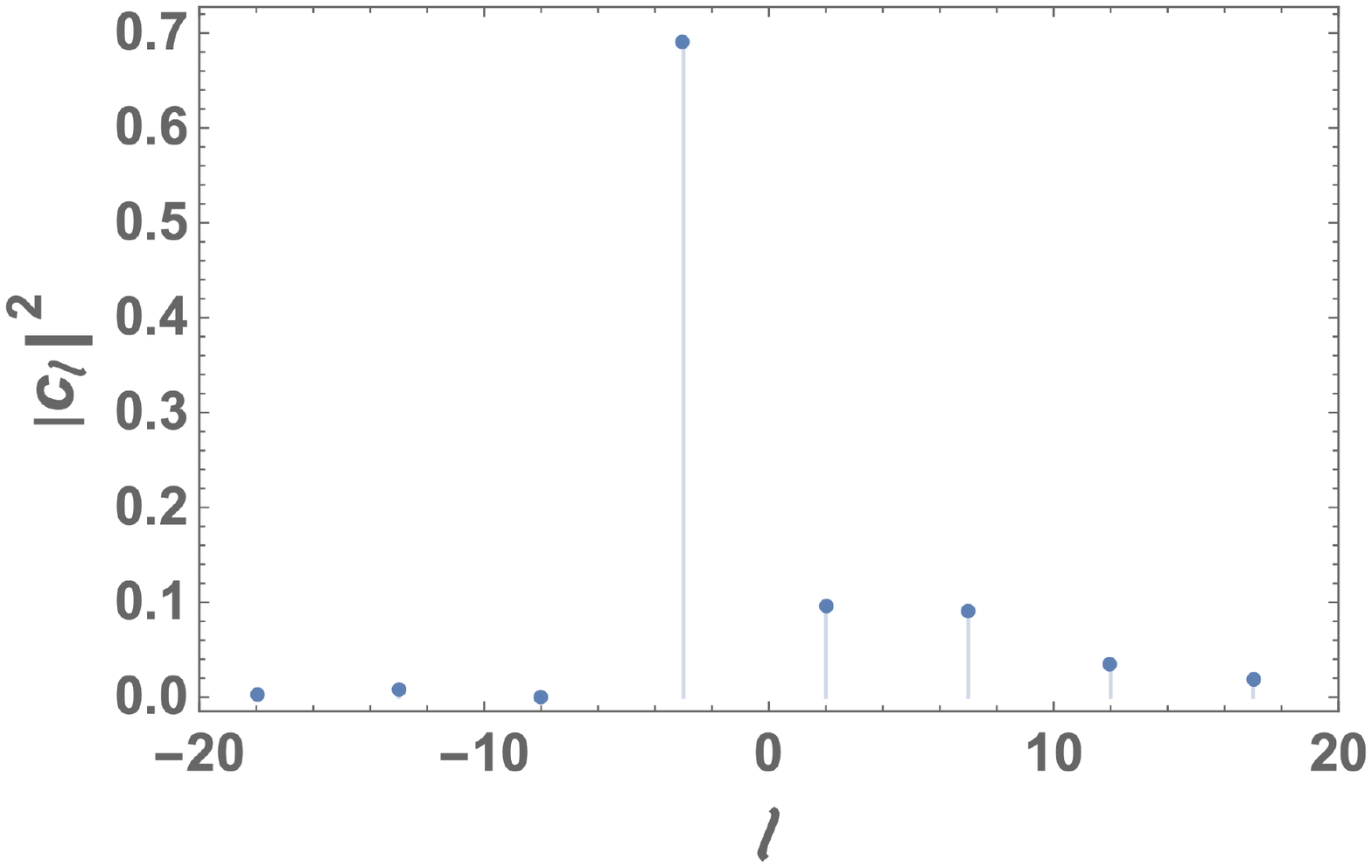}
\end{center}
\caption{OAM power spectrum arising from the azimuthal phase profile corresponding to the values of the parameters  $a=b=1$, $m=5$, $n_1=1/2$ and $n_2=n_3=4/3$.}\label{fig:Fig3}
\end{figure*}

This structure primarily affects the OAM spectrum, which includes only the components $(\ell - m) \pm k\, m$, with $k$ integer number (Fig.~\ref{fig:Fig3}), $\ell$ being the OAM index corresponding to the background helical mode. Specifically, in Fig.~\ref{fig:Fig3}, it has been reported the OAM power spectrum $|c_l|^2$ of the generated FFDH. In classical optics, the quantity $|c_l|^2$ is the fraction of the total power of the optical field component carrying an OAM proportional to $l$. In quantum optics, it is the probability that a photon in the beam carries an OAM of $\hbar l$. The actual values of $|c_l|^2$, as reported in Fig.~\ref{fig:Fig3}, have been determined numerically, by Fourier expanding the azimuthal phase factor reported in Eq.~(\ref{eq:normal}). The skew rays follow the paths dictated by $k_{\phi}$.
 
\section{Free-Form Azimuthal (FFA) SVAPs}
\begin{figure*}[h!]
\begin{center}
\includegraphics[width=10cm]{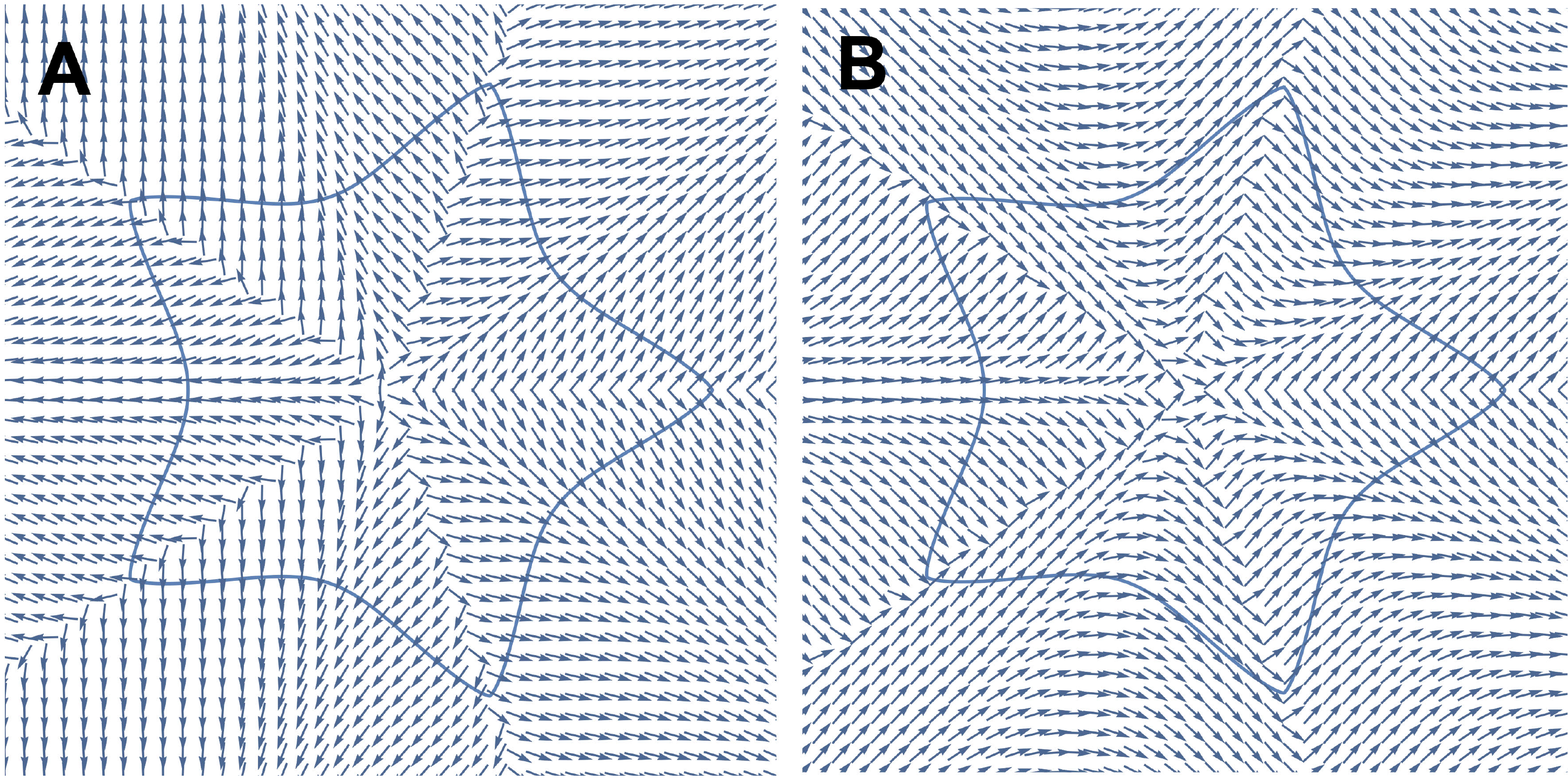}
\end{center}
\caption{Optic axis patterns deduced from Eq.~(\ref{eq:SuperSVAP}) for the values of the parameters $a=b=1$, $m=5$, $n_1=1/2$ and $n_2=n_3=4/3$. Inset {\bf A}: optic axis pattern for a SVAP imparting to an input beam the geometric phase $2\phi + 2 \Delta \Theta_m(\phi)$ (Fig.~\ref{fig:Fig1}~{\bf B}). Inset {\bf B}: optic axis pattern for a SVAP imparting the geometric phase $2 \Delta \Theta_m(\phi)$ (Fig.~\ref{fig:Fig1}~{\bf C}).\label{fig:Fig4}}
\end{figure*}

\begin{figure*}[h!]
\begin{center}
\includegraphics[width=10cm]{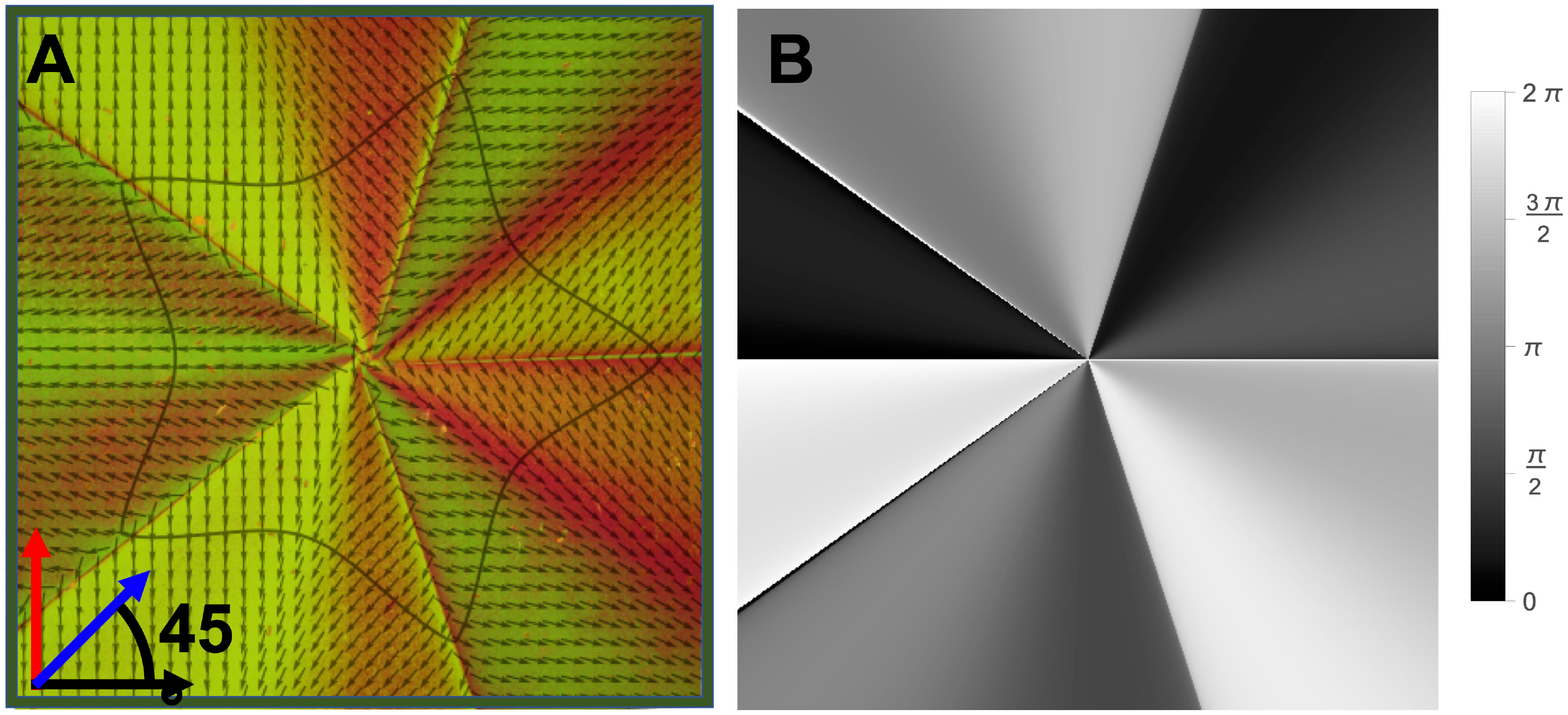}
\end{center}
\caption{Experimental observation of the optic axis distribution of the SVAP ($a=b=1$, $m=5$, $n_1=1/2$ and $n_2=n_3=4/3$). Inset {\bf A}: microscope image of the SVAP  between crossed polarizers + birefringent compensator plate at 45$^\circ$. This image unveils the optic axis pattern underlying the SVAP (Fig.~\ref{fig:Fig4}~{\bf A}), which is displayed in the image overlay. The image was recorded illuminating with white light the sample, sandwiched between crossed polarizers, and inserting, between the sample and the analyzer, a birefringent $\lambda$-compensator ($\lambda=550$~nm), having the optic axis rotated by 45$^\circ$. The arrows in the lower left corner sketch the axes orientations of the input linear polarizer (black arrow), the output analyzer (red arrow) and the $\lambda$-compensator (blue arrow). Inset {\bf B}: optical transverse phase profile associated to the optic axis pattern in inset {\bf A} -- the same as in Fig.~\ref{fig:Fig1}~{\bf B} -- here replicated for the sake of comparison.}\label{fig:Fig5}
\end{figure*}

Let's now focus the attention on the experimental methods for generating optical beams having the phase structure prescribed by Eq.~(\ref{eq:superphase}). To reshape a TEM$_{00}$ laser beam according to our wishes, we opted in favor of a properly tailored SVAP. The latter is a half-wave retardation plate in which the direction-angle of $\bar{\Theta}(r,\phi)$ of the optic-axis is spatially variant~\citep{piccirillo10, 10.1117/12.2078372, Aleman-Castaneda:19}. When a circularly-polarized input beam passes through the plate, it acquires a \textit{geometric phase} factor $e^{\pm{\rm i}2\bar{\Theta}(r,\phi)}$. The sign in the exponent depends on the handedness of the incident beam polarization $\bm{C}_\pm =\left(\bm{x}\pm {\rm i} \bm{y}\right)/\sqrt{2}$, which is reversed by the SVAP~\citep{Piccirillo:17}. For a comprehensive view of the mechanism underlying wavefront reshaping via Geometric or Pancharatnam-Berry Phase, we address the reader to Ref.~\citep{Piccirillo:17}. In essence, moulding the phase of a SVAP amounts to pattern the optic-axis so that its direction-angle is locally equal to half the prescribed optical phase. In order to fabricate a liquid-crystal SVAP for generating FFDH beams, the optic-axis angular distribution must be set to
\begin{equation}\label{eq:SuperSVAP}
\bar{\Theta}(r,\phi)=\frac{\psi(\phi)}{2}= \Theta(\phi; a,b,m,n_1,n_2,n_3).
\end{equation} 
In Fig.~\ref{fig:Fig4}, we show the optic-axis pattern of a SVAP corresponding to $\Theta(\phi; a=1,b=1,m=5,n_1=1/2,n_2=4/3,n_3=4/3)$ (inset {\bf A}) and, for comparison, the contribution to such pattern due to the modulation only (inset {\bf B}). In Fig.~\ref{fig:Fig5} ({\bf A}), it is shown a microscope image of the SVAP between crossed polarizers, with a birefringent $\lambda$-compensator inserted between the SVAP and the analyzer. The $\lambda$-compensator has a path difference of 550 nm and therefore introduces a $\pi$ retardation at that wavelength. The fast axis forms a 45$^\circ$ angle to the axis of the analyzer. When the compensator is put in, the sample changes its color depending on its orientation. The changes in color are based on optical interference. This method fully unveils the optic axis pattern underlying the SVAP (Fig.~\ref{fig:Fig4} ({\bf A})), because, differently from the simple crossed-polarizers method, it enables to distinguish between orthogonal orientations of the optic axis.

Though pure-phase holograms displayed on SLM could be used to create FFDH beams, fabricating optical devices based on Geometric Phase have proved to be not only the most performing choice, but also the most natural, since the unit normal distribution deduced from a generating curve is directly translated into an optic axis pattern. As an example, we have here chosen curves generated via superformula, to take advantage of a large variety of shapes grouped under the same equation. A similar method, however, can be applied to any other curve or family of curves. 

\section{Intensity profiles}
\begin{figure*}[h!]
\begin{center}
\includegraphics[width=10cm]{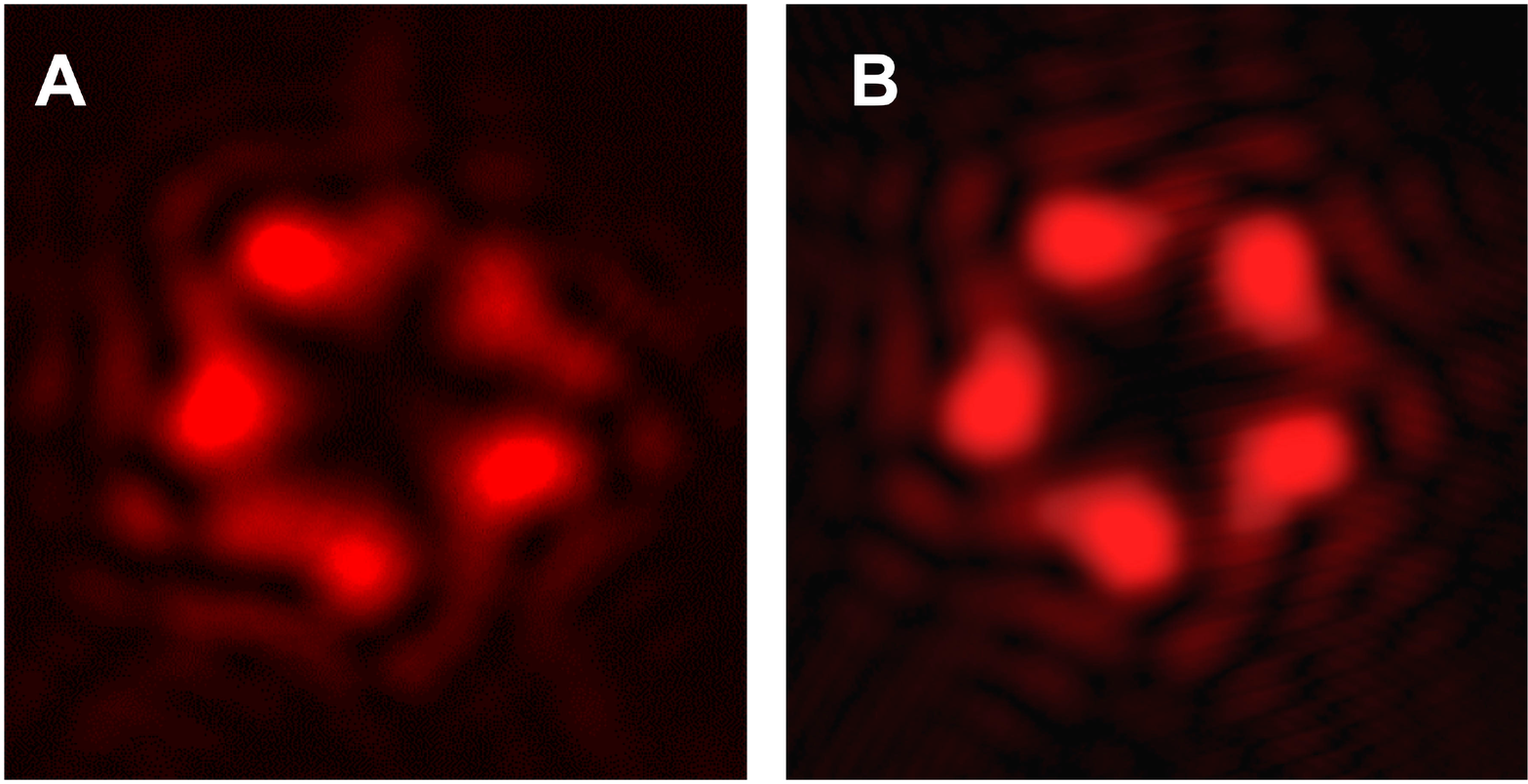}
\end{center}
\caption{Comparison between the experimental (inset {\bf A}) and theoretical (inset {\bf B}) intensity profiles of the beam generated through the SVAP having the optic axis pattern shown in Fig.~\ref{fig:Fig4}~{\bf A} at distance $z=1$~m, for the values  $a=b=1$, $m=5$, $n_1=1/2$ and $n_2=n_3=4/3$ of the curve parameters and for an input gaussian beam with plane wavefront and radius $w_0=(1.50 \pm 0.04)$~mm.}\label{fig:Fig6}
\end{figure*}

As above mentioned, by adding a periodical azimuthal phase modulation to the phase of a helical beam, the cylindrical symmetry typical of the intensity profile of a doughnut is broken. In fact, each photon at distance $r$ from the beam axis suffers a change in its azimuthal linear momentum $k_{\phi}$ that depends periodically on the orientation of the meridional plane it starts from. As $k_{\phi}$ has the same period as $\rho(\phi)$ in Eq.~(\ref{eq:rho}), the resulting transverse intensity profile turns to be periodic as well. What's more, the details of the  profile of $k_{\phi}$ are inherited from the azimuthal rate of change of the unit vector normal to the curve, therefore also the inflections of the intensity profiles will be inherited from the local curvature of the generating curve. This enables to set a one to one correspondence between the geometric properties of the generating curve and the transverse intensity profile of the beam, especially as far as is concerned the dark region. In Fig.~\ref{fig:Fig6}~{\bf A}, we show the intensity profile of the beam experimentally generated for the values of the parameters $a=b=1$, $m=5$, $n_1=1/2$ and $n_2=n_3=4/3$ at distance $z=1$~m from the SVAP, for a circularly polarized input TEM$_{00}$ gaussian mode with plane wavefront and radius $w_0=(1.50 \pm 0.04)$~mm. For comparison, in inset {\bf B} of Fig.~\ref{fig:Fig6}, it is shown the theoretical intensity profile predicted by calculating the Fresnel transform of the optical field
\begin{equation}\label{eq:Fresnel}
E_0 \, {\rm e}^{-\frac{x^2+y^2}{w_0^2} + 2\,{\rm i} \, \Theta(\phi; 1, 1, 5, 1/2, 4/3, 4/3)},
\end{equation} 
for the same values of the parameters. The faint striped structure surrounding the core profile originates by diffraction from the abrupt azimuthal changes in the transverse phase profile shown in Fig.~\ref{fig:Fig5}~{\bf B}.

\section{Concluding remarks}
We have shown the possibility to generate dark hollow beams with a large variety of intensity landscapes by using a single azimuthal phase factor without passing through numerical methods for the optical field encryption. The method is based on a geometric approach in which the intensity profile around the beam axis is supposed to imitate the shape of a selected closed curve. Also the OAM spectrum is affected by the shape of the generating curve. If the generating curve has an $m$-fold rotational symmetry, the OAM spectrum will include only components with multiple of $m$ within a global shift determined by the OAM index of the unperturbed helical mode. Liquid-crystals SVAPs turn to be the most natural choice to implement such method, since the unit vector normal to the generating curve come to be copied over the axis pattern. Applications of FFA SVAPs can be easily devised, in particular, for manipulating non-spherical objects trapped by optical tweezers -- as unwanted rotations of micro-objects could be avoided -- as well as for increasing contrast in optical coronagraphy -- as properly tailored dark-hollow beams with line singularities along radial directions could be exploited to split the intensity distribution around the optical axis.





\section*{Conflict of Interest Statement}

The authors declare that the research was conducted in the absence of any commercial or financial relationships that could be construed as a potential conflict of interest.

\section*{Author Contributions}

All the authors contributed to develop the method for designing the Free-Form SVAPs introduced in the paper. BP fabricated and tested the SVAPs. All the authors contributed to writing the paper.

\section*{Funding}
This work was supported by the University of Naples Research Funding Program (DR n. 3425-10062015) and by the European Research Council (ERC), under grant no. 694683 (PHOSPhOR).

\section*{Acknowledgments}
The authors thank prof. L. Marrucci from the Department of Physics "E. Pancini" of University of Naples for useful discussions.

\section*{Figure captions}

%

\end{document}